# Asteroid body-fixed hovering using nonideal solar sails

Xiang-yuan Zeng, Fang-hua Jiang, Jun-feng Li
School of Aerospace Engineering, Tsinghua University, Beijing 100084, China;
*Corresponding Email: jiangfh@tsinghua.edu.cn*

**Abstract:** Asteroid body-fixed hovering problem using nonideal solar sail models in a compact form with controllable sail area is investigated in this paper. The nonlinear dynamic equations for the hovering problem are constructed for a spherically symmetric asteroid. The feasible region for the body-fixed hovering is solved from the above equations by using a shooting method. The effect of the sail models, including the ideal, optical, parametric and solar photon thrust, on the feasible region is studied through numerical simulations. The influence of the asteroid spinning rate and the sail area-to-mass ratio on the feasible region is discussed in a parametric way. The required sail orientations and their corresponding variable lightness numbers are given for different hovering radii to identify the feasibility of the body-fixed hovering. An attractive mission scenario is introduced to enhance the advantage of the solar sail hovering mission.

## 1 INTRODUCTION

CLOSE proximity missions to hazardous asteroids (Scheeres 2004) have been frequently investigated to be as a precursor to some mitigation strategy or a controlled landing. According to previous studies, there are a number of possible options for explorations in the vicinity of the asteroid, including Sun synchronous orbit (Morrow et al. 2001), retrograde orbit (Broschart & Scheeres 2005) and heliostationary flight (Morrow et al. 2002) in which a spacecraft is placed in the system libration point (Baoyin & McInnes 2005). Besides the above methods, the spacecraft could also maintain a required fixed position relative to the rotating asteroid referred to be as the 'body-fixed hovering'. It is an effective way for an asteroid human landing or a sample return mission which has been successfully implemented by Hayabusa (Scheeres 2004). If the mission requires prolonged observation of a specific area locating away from the synchronous orbit, the thrust must accommodate both the gravitational and centrifugal forces. Thus, the extended hovering period will highly depend on the onboard supplies of fuel for chemical or continuous low-thrust spacecraft. Compared to the conventional spacecraft, the inherent capabilities of solar sailing without fuel consumptions make them well suited for asteroid explorations.

Close asteroid orbital dynamics is quite challenging and complex due to their irregular shape and rotation (Hu & Scheeres 2008; Li et al. 2013). Additionally, the solar radiation pressure (SRP) becomes a significant perturbing force in the vincinity of those small sized asteroids (Scheeres 1999). Another advantage of solar sailing is to utilize the SRP force as an active control. The first detailed analysis about sail operations at asteroids was made by Morrow and Scheeres (2001). Sawai (2002) and Broschart and Scheeres (2005) investigated the body-fixed hovering with conventional propulsion systems. Zhang (2013) extended such hovering from satellite-to-asteroid to the case of two satellites. However, solar sail asteroid body-fixed hovering was not addressed before Williams' work (2009). In his work (Williams & Abate 2009), an ideally reflecting sail model associated with a sail efficiency factor (to reflect the difference between a true sail and an ideal one) was adopted whose corresponding SRP force is normal to the sail surface. Nonideal sails have not yet been discussed regarding asteroid body-fixed hovering although some heliostationary flight has ever been presented by Morrow (2002) and Jorba (2012).

Solar sail has been seriously considered as an alternative propulsion system since the comet Halley rendezvous mission. A number of demonstrative missions (McInnes 1999; Baoyin & McInnes 2006) have been investigated along with their corresponding practical experiments. Some dramatic mission concepts involving non-Keplerian orbits (Gong et al. 2007 & 2009a; Vulpetti 1997) have been proposed by using solar sailing. The successful flying of IKAROS and NanoSail-D2 has gained a lot of interest from the space community and laid a first stone for further sail missions (Gong et al. 2011). A concept of furlable solar sail was proposed by Williams (2009) to generalize the body-fixed hovering region. Compared to a fixed-area solar sail with two variable attitude angles, the essence of a furlable sail is to separate the maximum SRP force magnitude as an independent control variable. Such a performance can be also implemented with a variable reflectivity sail film which has been partially demonstrated on IKAROS to control the sail attitude.

In this paper a compact form of nonideal sails (Mengali & Quarta 2007) with controllable sail area is adopted to accomplish the asteroid body-fixed hovering. A comparison is made to quantify the influence of the four different (including optical-, parametric-, and solar photon thrust) sail models on the hovering mission. For the target asteroid, a spherically symmetric model is applied to be as an estimation of the first step. The asteroid model can be relaxed and extended in future studies. The analysis presented here complement the studies made by Williams (2009) and extend to the scenarios with realistic sails. Moreover, the effect of the asteroid rotation and the sail area-to-mass ratio on the body-fixed hovering region is also illustrated via numerical simulations. Finally, the sail control profiles corresponding to different hovering radii are presented to identify the feasibility of the body-fixed hovering by using solar sailing.

## 2 BODY-FIXED HOVERING FORMULATION

### 2.1 Equations of Motion

In this analysis, a two-body gravitational model is adopted to describe the dynamics of the spacecraft near an asteroid. The vector dynamical equation for a solar sail in the uniformly rotating body-fixed coordinate frame *oxyz* (Scheeres et al. 1998) can be written as

$$\frac{d^2\bm{r}}{dt^2} + 2\bm{\omega}\times\frac{d\bm{r}}{dt} + \bm{\omega}\times(\bm{\omega}\times\bm{r}) = \bm{a}_{\text{SRP}} - \frac{\partial U(\bm{r})}{\partial \bm{r}} \qquad (1)$$

where $\bm{r}$ is the position vector from the asteroid center of mass to the sailcraft, $\bm{\omega}$ is the rotational angular velocity vector of the asteroid with respect to the inertial reference frame *IXYZ*, $U(\bm{r})$ is the gravitational potential of the asteroid and $\bm{a}_{\text{SRP}}$ is the non-conservative solar radiation pressure (SRP) acceleration. The coordinate system *IXYZ* centered on the asteroid is given in Fig. 1. The +*IZ* axis is along the direction of the asteroid angular velocity, the +*IX* axis is along the anti-solar direction and in the asteroid equatorial plane, and +*IY* axis is also in the asteroid equatorial plane making up an orthogonal right-handed triad. Strictly speaking, the inertial frame *IXYZ* is a near non-rotating coordinate system due to the conic motion of the asteroid. However, compared to the asteroid rotation period on the order of hours to days at most, the rotation of the frame *IXYZ* is on the order of thousand days beyond 2 AU away from the Sun. Thus, for those asteroids in the main belt, the frame *IXYZ* can be treated as an inertial reference frame.

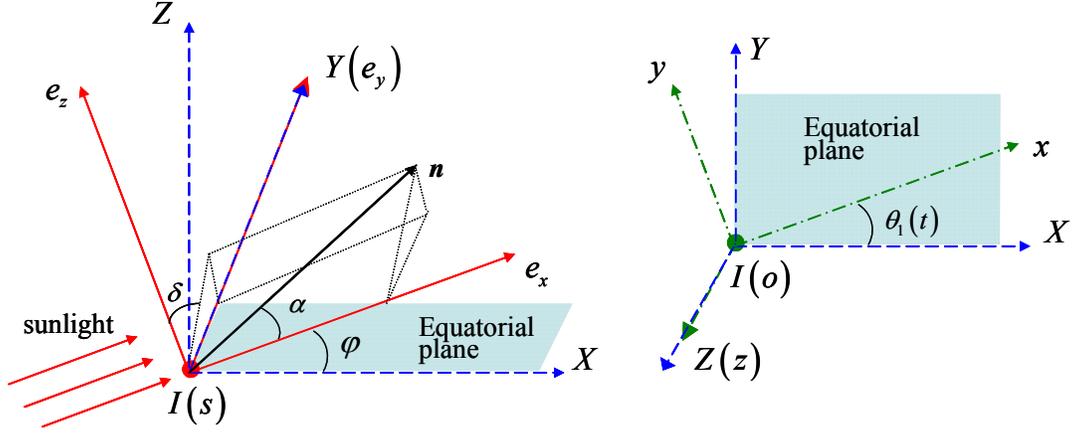

**Fig. 1** Orbital reference frames and sail attitude angles.

The body-fixed frame *oxyz* coincides with the frame *IXYZ* at the initial time and the transformation matrix from *oxyz* to *IXYZ* is

$$C_1(t) = \begin{bmatrix} \cos\theta_1 & -\sin\theta_1 & 0 \\ \sin\theta_1 & \cos\theta_1 & 0 \\ 0 & 0 & 1 \end{bmatrix} \quad (2)$$

where the angle $\theta_1(t) = \omega t$. In order to express the SRP force, an incident light coordinate system $se_x e_y e_z$ shown in Fig. 1 is established where the $+se_x$ axis is along the sunlight direction. The axis $+se_y$ coincides with the $+IY$ axis and $se_z$ completes the right-handed frame. In this frame, the unit vector $\mathbf{s}$ directed from the sun to the asteroid is always $[1, 0, 0]^T$, which is the same as the unit vector of the axis $+se_x$. If there is a solar latitude angle $\varphi$ between the sunlight and the asteroid equatorial plane, the transformation matrix from *IXYZ* to $se_x e_y e_z$ is

$$C_2 = \begin{bmatrix} \cos\varphi & 0 & \sin\varphi \\ 0 & 1 & 0 \\ -\sin\varphi & 0 & \cos\varphi \end{bmatrix}, \quad \varphi \in \left[-\frac{\pi}{2}, \frac{\pi}{2}\right] \quad (3)$$

Seen from Fig. 1, if axis $+se_x$ is along axis $+IZ$ corresponding to $\varphi = \pi/2$, the Sun locates at the south pole of the asteroid. If $\varphi = 0$ the sunlight is in the asteroid equatorial plane.

### 2.2 Solar Sail Force Model

A unified, compact form of solar sails with fixed sail area has been presented by Mengali and Quarta (2007) to accomplish the advanced heliostationary missions. Compared to the ideal sail with a perfectly flat reflective surface, the optical model takes the effect of reflection, absorption and reradiation into account. The parametric model considers the billowing of the sail. Additionally, the solar photon thrust (SPT, detailed by Guerman et al. 2009) is also included in the model. The sail acceleration with variable sail area can be written as

$$\mathbf{a}^s(t) = \frac{\beta(t)}{2} \cdot \frac{\mu_{sun}}{R_{AU}^2} \cdot \cos^{(p-q)}\alpha \cdot \left[(1-q)b_1 \cdot \mathbf{e}_x^s + \left(qb_1 + b_2\cos^{(3q+1)}\alpha + b_3\cos^{2q}\alpha\right) \cdot \mathbf{n}^s\right] \quad (4)$$

where the superscript 's' indicates that the vector is expressed in the $se_x e_y e_z$ frame. In the above equation, $\mu_{sun}$ is the solar gravitational constant ($1.3271244 \times 10^{20}$ m$^3$ s$^{-2}$) and $R_{AU}$ is the Sun-asteroid heliocentric distance in the unit of AU (1 AU $\approx 1.496 \times 10^{11}$ m). The coefficients [$p$, $q$, $b_1$, $b_2$, $b_3$] corresponding to different sail models are specified in Table 1 (but see discussions by Mengali & Quarta 2007).

**Table 1** Solar sail force model coefficients

| $p$ | $q$ | $b_1$ | $b_2$ | $b_3$ |
| --- | --- | --- | --- | --- |

| | | | | | |
|---|---|---|---|---|---|
| Ideal | 1 | 0 | 0 | 2 | 0 |
| Optical | 1 | 0 | 0.1728 | 1.6544 | −0.0109 |
| Parametric | 1 | 1 | −0.5885 | −0.1598 | 2.5646 |
| SPT | 0 | 0 | 0 | 2 | 0 |

The sail cone angle $\alpha$ shown in Fig. 1 is defined as the angle between the sail normal vector $\boldsymbol{n}$ and the incident light $\boldsymbol{e}_x$ ($\boldsymbol{e}_x^s = [1, 0, 0]^T$) for both ideal and optical models. However, for the parametric and SPT sails, $\alpha$ is the angle between the SRP force and the vector $\boldsymbol{e}_x$. The sail orientation can be explicitly expressed as

$$\boldsymbol{n}^s = \begin{bmatrix} \cos\alpha \\ \sin\alpha \sin\delta \\ \sin\alpha \cos\delta \end{bmatrix}, \quad \begin{cases} \alpha \in [0, \pi/2] \\ \delta \in [0, 2\pi) \end{cases} \tag{5}$$

where $\delta$ is the clock angle shown in Fig. 1, defined as the angle between the projected line of the incident light onto the plane $se_y e_z$ and axis $+se_z$. For an ideal sail, Eq. (4) becomes

$$\boldsymbol{a}^s(t) = \beta(t) \cdot \frac{\mu_{sun}}{R_{AU}^2} \cdot \cos^2\alpha \cdot \boldsymbol{n}^s \tag{6}$$

where the sail lightness number $\beta(t)$ is the ratio of the SRP acceleration to the local solar gravitational acceleration, which only depends on the sail area-to-mass ratio $\sigma(t)$

$$\beta(t) = \frac{\sigma_{SL}}{\sigma(t)} = \frac{\sigma_{SL}}{m/A(t)}, \quad \beta \leq \beta_{max} \tag{7}$$

where the critical sail loading parameter $\sigma_{SL}$ is a constant whose value is approximately 1.53 g m$^{-2}$, $m$ is the total mass of the sailcraft and $A(t)$ is the effective reflective surface of the sail. Here, $\beta_{max}$ is the maximum available lightness number which is key design parameter for a mission.

### 2.3 Body-Fixed Hovering

The body-fixed hovering leads to a fixed equilibrium point in the frame $oxyz$ at a desired position. From Eq. (1), the enabling SRP acceleration should be

$$\boldsymbol{a}_{SRP} = \frac{\partial U(\boldsymbol{r})}{\partial \boldsymbol{r}} + \boldsymbol{\omega} \times (\boldsymbol{\omega} \times \boldsymbol{r}) = \begin{bmatrix} U_x \\ U_y \\ U_z \end{bmatrix} - \begin{bmatrix} \omega^2 & 0 & 0 \\ 0 & \omega^2 & 0 \\ 0 & 0 & 0 \end{bmatrix} \begin{bmatrix} x \\ y \\ z \end{bmatrix} \tag{8}$$

The desired position can be expressed by the latitude angle $\lambda \in [-\pi/2, \pi/2]$ and the longitude angle $\theta_0 \in [0, 2\pi]$ as

$$\boldsymbol{r} = \begin{bmatrix} x \\ y \\ z \end{bmatrix} = r \begin{bmatrix} \cos\lambda \cos\theta_0 \\ \cos\lambda \sin\theta_0 \\ \sin\lambda \end{bmatrix} \tag{9}$$

where $r$ is the magnitude of the position vector. It is assumed here that the asteroid is spherically symmetric. Thus, the gravitational acceleration exerted on the sailcraft is

$$\frac{\partial U(\boldsymbol{r})}{\partial \boldsymbol{r}} = \begin{bmatrix} U_x \\ U_y \\ U_z \end{bmatrix} = \frac{\mu_{ast}}{r^3} \boldsymbol{r} = \frac{\mu_{ast}}{r^2} \begin{bmatrix} \cos\lambda \cos\theta_0 \\ \cos\lambda \sin\theta_0 \\ \sin\lambda \end{bmatrix} \tag{10}$$

where $\mu_{ast}$ is the gravitational constant of the asteroid. Substituting Eq. (9) and Eq. (10) into Eq. (8), the sail acceleration is given as

$$\boldsymbol{a}_{SRP} = \left[ \left(\frac{\mu_{ast}}{r^2} - \omega^2 r\right) \cos\lambda \cos\theta_0, \; \left(\frac{\mu_{ast}}{r^2} - \omega^2 r\right) \cos\lambda \sin\theta_0, \; \frac{\mu_{ast}}{r^2} \sin\lambda \right]^T \tag{11}$$

A constraint of the solar sail is that the SRP force can be only produced in the anti-solar hemisphere. Therefore, the sail acceleration must satisfy

$$\left(\boldsymbol{a}_{\text{SRP}}^{s}\right)^{\text{T}} \cdot \boldsymbol{e}_{x}^{s} = \left(C_{2} \cdot C_{1}(t) \cdot \boldsymbol{a}_{\text{SRP}}\right)^{\text{T}} \cdot \boldsymbol{e}_{x}^{s} \geq 0 \tag{12}$$

Substituting Eq. (2) and Eq. (3) into Eq. (12), one can obtain

$$\left(\boldsymbol{a}_{\text{SRP}}^{s}\right)^{\text{T}} \cdot \boldsymbol{e}_{x}^{s} = \left(\frac{\mu_{\text{ast}}}{r^{2}} - \omega^{2} r\right) \cos\lambda \cos\varphi \cos\theta + \frac{\mu_{\text{ast}}}{r^{2}} \sin\lambda \sin\varphi \geq 0 \tag{13}$$

where $\theta(t) = \theta_0 + \theta_1(t) = \theta_0 + \omega t$. Since the body-fixed hovering of a sailcraft covers the whole period of the asteroid rotation, the angle $\theta(t)$ takes all values from 0 to $2\pi$. In order to guarantee the value of Eq. (13) to be always positive, the right second term $\mu_{\text{ast}} \sin\lambda \sin\varphi / r^2$ should be always positive in that $\cos\theta \in [-1, 1]$. It indicates that the angles $\lambda$ and $\varphi$ must be of the same sign. According to the definitions of these two angels, the corresponding situation is that the sailcraft and the Sun must lie in different sides of the asteroid equatorial plane.

To accomplish the body-fixed hovering, the required SRP acceleration in Eq. (11) should be the same as that provided in Eq. (4):

$$\boldsymbol{\Pi} = \boldsymbol{a}_{\text{SRP}}^{s} - \boldsymbol{a}^{s}(t) = C_{2} \cdot C_{1}(t) \cdot \boldsymbol{a}_{\text{SRP}} - \boldsymbol{a}^{s}(t) = \boldsymbol{0} \tag{14}$$

There are three control variables ($\beta$, $\alpha$, $\delta$) corresponding to the above three dimensional nonlinear equations. It seems impossible to obtain analytical solutions but could be solved numerically. For a specified position as expressed in Eq. (9), if there are a set of time-variant ($\beta$, $\alpha$, $\delta$) making Eq. (14) zero when $\theta_1$ takes all values of 0 to $2\pi$, the hovering point is feasible and vise versa. The nonlinear equations can be solved using the Matlab's '*fsolve*' function with a default method of '*dogleg*'. In order to improve the calculation efficiency, a program of MinPack-1 translated into C++ language is adopted to solving the nonlinear equations. In all simulations, the tolerance of Eq. (14) is satisfied to be better than $10^{-9}$. Since Eq. (14) is only three dimensional without integrations, it is not sensitive to the initial values.

During the simulations per running when $\theta_1$ takes all values of 0 to $2\pi$, there are some "bad" cases needed to be verified corresponding to feasible solutions (i.e. $0 < \beta \leq \beta_{\text{max}}$, $\|\alpha\| \leq \pi/2$). Specifically, the control attitude angles must be in their feasible domains. There have been two cases arisen in our simulation process which can be transformed into feasible solutions. The two cases and their equivalent expressions are

$$\begin{cases} \alpha \in (-\pi/2, 0] \\ \delta > 0 \end{cases} \Rightarrow \begin{cases} \alpha = \|\alpha\| \\ \delta = \text{mod}(\pi + \delta, 2\pi) \end{cases}, \quad \begin{cases} \alpha \in [0, \pi/2) \\ \delta \in [-2\pi, 0] \end{cases} \Rightarrow \begin{cases} \alpha = \alpha \\ \delta = \text{mod}(2\pi + \delta, 2\pi) \end{cases} \tag{15}$$

where the function 'mod' is the modulus after division.

## 3 CASE STUDY

The effect of sail coefficients, the asteroid rotating angular velocity and the sail area-to-mass ratio on the feasible region of the body-fixed hovering will be examined in this section. The influence of the hovering radius on the sail control history is investigated through numerical simulations. The main parameters of the asteroid considered here are the same as given by Williams (2009). The heliocentric orbit of the asteroid is assumed to be circular at 2.7 AU. Its diameter is 1.0 km with a density of $2.4 \times 10^3$ kg m$^{-3}$ and its rotational period is 9.0 hours with spin axis aligned with axis $+oz$. The highest solar lightness number is 0.153 corresponding to the area-to-mass ratio of 10 g m$^{-2}$. The solar latitude angle is set to be 60 deg for such a main-belt asteroid. Except special explanations, all simulations in this section are in the above parameters.

### 3.1 Effect of Solar Sail Force Coefficients

Figure 2 shows the feasible body-fixed hovering regions with four different sail models whose coefficients are already given in Table 1. Since the asteroid is assumed to be spherically symmetric, feasible regions for $\theta_0$ from 0 to $2\pi$ should be the same. Those feasible regions locate at the asteroid's northern hemisphere since the Sun is below its equatorial plane ($\varphi = 60$ deg). The region of the SPT sail is largest while the smallest is parametric. Feasible regions of the optical and parametric sails are nearly the same although their model coefficients are totally different.

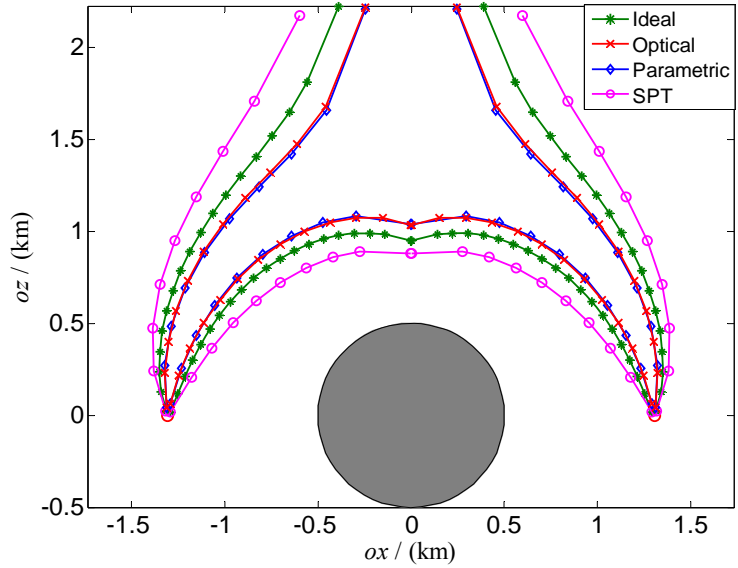

**Fig. 2** Feasible regions of four different sail models.

For those four sail models, all regions start from the corresponding synchronous points in the equatorial plane at a radius of $r_{syn}$ ($r_{syn} = \sqrt[3]{\mu_{ast}/\omega^2}$, here $r_{syn} \approx 1.31$ km). When the sailcraft reaches the asteroid's northern pole, all SRP force is used to counterbalance the asteroid gravitational force. For an ideal sail, it is easy to identify that the closest hovering radius is

$$r_{min} = \sqrt{\frac{\mu_{ast}}{\beta_{max} \cdot (\mu_{sun}/R_{AU}^2) \cdot \sin^2 \varphi}} \quad (16)$$

whose value in our case is approximately 0.95 km. The value of $r_{min}$ for optical and parametric sails is 1.03 km while that is 0.88 km for the SPT sail. As a rough estimation, an efficiency factor of 0.85 can be added to Eq. (6) to approximate the optical and parametric models here (see results of Williams 2009). The shadow effect of the asteroid from the Sun is neglected which reduces the effective hovering region for such a spherical model. For other irregular shaped asteroids, such an effect needs to be discussed in detail.

### 3.2 Asteroid Rotational Effects

Figure 3 illustrates the effect of asteroid spinning rate on the hovering regions with the asteroid at the lower center. Two spin angular velocities are considered, i.e., 9 hours and 15 hours (which is arbitrarily selected longer than 9 hours). Only the ideal sail model is adopted here and that's why the minimum anti-sun pole radius $r_{min}$ is the same at 0.95 km. For the hypothetical case of slow rotating the synchronous orbit radius is approximately 1.84 km. There is overlapping area between the two regions when the hovering latitude angle $\lambda$ is greater than $0.205\pi$. The feasible hovering region of the lower $\omega$ is a little larger than the higher case. It indicates that there are more options of hovering positions (but with relatively farther hovering radius for the same latitude) about slow rotating asteroids compared to those of similar physical properties and orbital parameters.

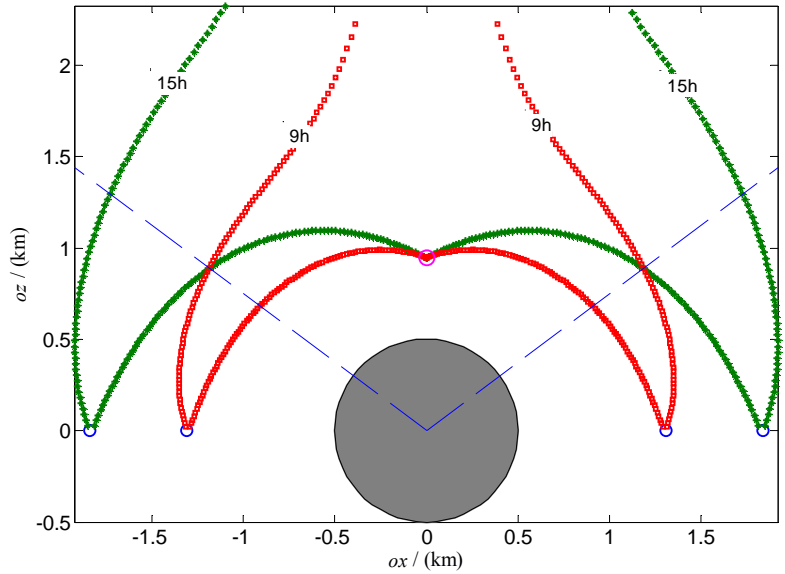

**Fig. 3** Effect of asteroid rotation on the feasible hovering region.

### 3.3 Effect of the Area-to-Mass Ratio

In this section, the effect of the sail area-to-mass ratio $\sigma(t)$ on the hovering radius will be investigated. According to recent studies, a characteristic acceleration on the order of 0.5 mm s$^{-2}$ can accomplish near-term sail missions while a relatively mid-term 160-m square sail has been envisaged by NASA for the Solar Polar Imager (SPI) mission (Mengali & Quarta 2009). The characteristic acceleration $a_c$ is defined as the maximum sail acceleration at 1 AU when the sail normal direction is along the sunlight direction. Thus, the lower boundary value of $\sigma_{min}$ is 10 g m$^{-2}$ corresponding to a near-term sail with $a_c = 0.91$ mm s$^{-2}$. The upper boundary of $\sigma_{min}$ is 4 g m$^{-2}$ whose $a_c$ is approximately 2.27 mm s$^{-2}$. For a sail with variable sail area, the minimum area-to-mass ratio $\sigma_{min}$ corresponds to the highest sail lightness number based on Eq. (7). The investigated values of $\sigma_{min}$ ($\beta_{max}$, $a_{cmax}$) have been given in Table 2 along with their corresponding minimum hovering radii. These four values are enough to illustrate the influence of $\sigma_{min}$ on the feasible hovering regions. Additionally, the minimum hovering radius for $\sigma_{min} = 4$ g m$^{-2}$ is 0.599 km which is only 100 m away from the asteroid surface. There is no need to be closer for an observation mission for such a fictitious asteroid.

**Table 2** Minimum hovering radius for different values of the area-to-mass ratio

| $\sigma_{min}$ / (g m$^{-2}$) | 4 | 6 | 8 | 10 |
|---|---|---|---|---|
| $\beta_{max}$ | 0.3825 | 0.2550 | 0.1913 | 0.1530 |
| $a_{cmax}$ / (mm s$^{-2}$) | 2.2682 | 1.5122 | 1.1344 | 0.9073 |
| $r_{min}$ / (km) | 0.599 | 0.734 | 0.848 | 0.947 |

Figure 4 shows the variation of the hovering radius with respect to the hovering latitude in terms of each area-to-mass ratio. The curves above the synchronous orbit on 1.31 km are the outer boundaries of each feasible region while below curves are inner boundaries. The body-fixed hovering radius is obtained with the variation of the hovering latitude in a step of $0.002\pi$. It is easy to know that the feasible region of $\sigma_{min} = 4$ g m$^{-2}$ is larger than the other three cases due to its higher characteristic acceleration. The biggest gap of hovering radius between different values of $\sigma_{min}$ in Fig. 4 occurs at the asteroid polar region in a value of 0.35 km from Table 2.

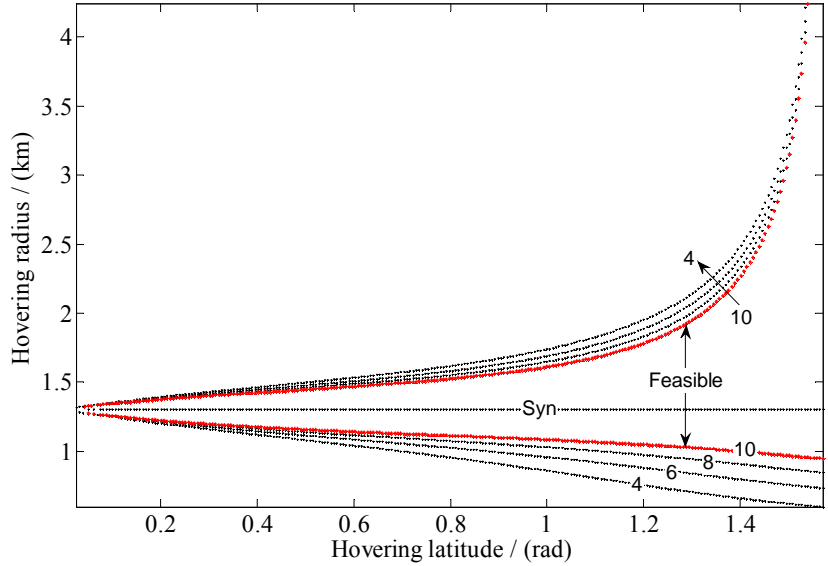

**Fig. 4   Relation between hovering radius and latitude for different sail area-to-mass ratios**

The body-fixed hovering for different latitudes with respect to the asteroid is feasible by using solar sailing. When the hovering latitude is specified, the increase of $\sigma_{min}$ in the above scenarios can increase a little of the hovering boundaries, especially for the areas away from the polar region. For example, if the hovering latitude $\lambda$ is 0.264$\pi$, the inner boundary for $\sigma_{min}$ = 4 g m$^{-2}$ is 0.94 km while 1.11 km for $\sigma_{min}$ = 10 g m$^{-2}$. The loss of 170 m hovering height for $\sigma_{min}$ = 10 g m$^{-2}$ makes its system mass 2.5 times than the case of $\sigma_{min}$ = 4 g m$^{-2}$. It indicates that there is a tradeoff between the hovering radius and the system payload mass. Taking the 160-m square sail of the demonstrated SPI mission as an example, the payload mass for $\sigma_{min}$ = 10 g m$^{-2}$ is 256 kg. On the contrary, the payload mass for $\sigma_{min}$ = 4 g m$^{-2}$ is only 102 kg. Therefore, it is preferable to take more payload mass for low-performance sails to accomplish the body-fixed hovering mission.

### 3.4   Effect of the Hovering Radius

The effect of the hovering radius on the sail control profile will be examined in this section. Without loss of generality, the hovering latitude is set to be $\pi/4$ and the minimum sail area-to-mass ratio is 10 g m$^{-2}$. The boundary values for this hovering latitude are 1116 m and 1522 m, respectively. Three hovering radii are investigated, i.e., 1156 m, 1306 m and 1456 m where 1306 m is the synchronous radius and the two others in the feasible region are 150 m away from the synchronous orbit. Since the incident light frame *IXYZ* is assumed to be non-rotating, the sail control profile for the body-fixed hovering is symmetrical with respect to the *IXZ* plane. Let's assume there are $2n + 1$ discrete points of the control profile between 0 to $2\pi$ where $n$ is an integer. In the current simulations, $n$ is 500 corresponding to the calculating step of $0.002\pi$ for the asteroid rotation. Then the control variables after $\pi$ can be obtained as

$$\begin{cases} \beta(n+1) = \beta(n-1) \\ \alpha(n+1) = \alpha(n-1) \\ \delta(n+1) = 2\pi - \delta(n-1) \end{cases} \quad (17)$$

The sail control profile in the first half period is shown in Fig. 5 for each case, including the sail lightness number and the two sail attitude angles. To distinguish the sail orientations in the bottom plotting of Fig. 5, the clock angle for the outward case (1456 m) is given as a segmented figure. For all these three cases, the peak of the lightness number does not exceed the maximum value of 0.153. For both the inward (1156 m) and outward cases, the lightness number and the sail attitude angles are time-variant to fulfill the requirement of the non-synchronous body-fixed hovering.

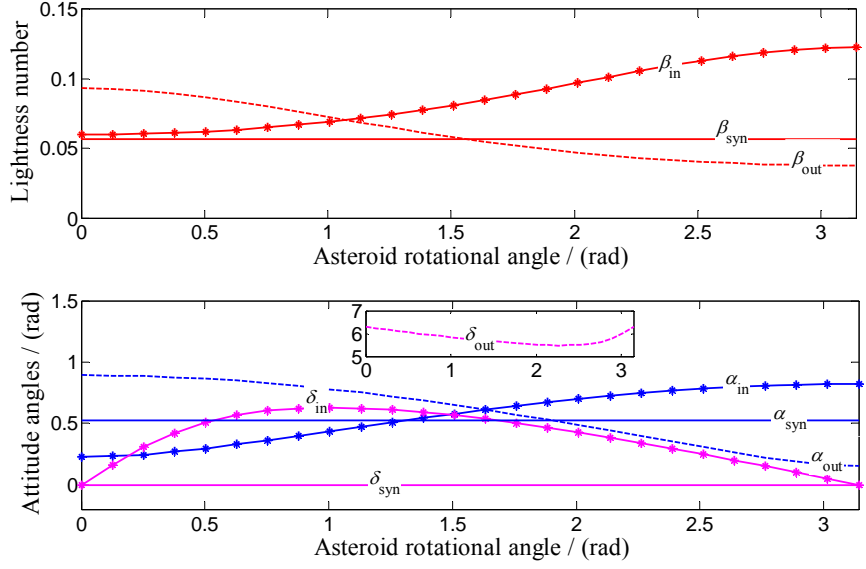

**Fig. 5**   Solar sail control profile in the first half period for different hovering radii.

It is interesting that the clock angle is always zero for the synchronous case (1306 m with $\lambda = \pi/4$) while the control variables $\beta$ and $\alpha$ are constant. It indicates that the required sail acceleration is only provided to counterbalance the asteroid gravitational acceleration along the axis $IZ$. Such a condition can be explicitly deduced for an ideal sail from section II. For a body-fixed hovering position at a synchronous height out of the equatorial plane, the required sail acceleration in Eq. (11) is simplified into

$$\boldsymbol{a}_{\mathrm{syn}} = \left[ 0, \ 0, \ \frac{\mu_{\mathrm{ast}}}{r_{\mathrm{syn}}^2} \sin \lambda \right]^{\mathrm{T}} \qquad (18)$$

Substituting Eq. (5) into Eq. (6) with $\delta = 0$, one can obtain

$$\boldsymbol{a}^s(t) = \beta(t) \cdot \frac{\mu_{\mathrm{sun}}}{R_{\mathrm{AU}}^2} \cdot \cos^2 \alpha \cdot \begin{bmatrix} \cos \alpha \\ 0 \\ \sin \alpha \end{bmatrix} \qquad (19)$$

In order to guarantee the feasibility of the body-fixed hovering, the provided sail acceleration of Eq. (19) should be equal to that required of Eq. (18). Therefore, the sail cone angle must satisfy $\alpha = \pi/2 - \varphi$ to make sure that the sail acceleration is along the $+IZ$ direction. In such a case, the required lightness number can be obtained as

$$\beta_{\mathrm{syn}} = \frac{\mu_{\mathrm{ast}}}{\mu_{\mathrm{sun}}} \cdot \frac{R_{\mathrm{AU}}^2}{r_{\mathrm{syn}}^2} \cdot \frac{\sin \lambda}{\sin^2 \varphi} \qquad (20)$$

For the above case, its sail lightness number can be calculated from Eq. (20) as 0.057 which is consistent with the value shown in Fig. 5. It is a good property that the asteroid body-fixed hovering can be accomplished by using solar sailing with constant control variables at the synchronous radius out of the equatorial plane. Such a property would be very attractive for future solar sailing body-fixed hovering missions. First, the self-rotating period of the asteroids is different from each other due to the large number of asteroids. Second, the sail attitude control for a large thin film is very challenging and complicated (Wie & Murphy 2007; Gong et al. 2009b). Therefore, for specified hovering latitudes out of the asteroid equatorial plane, the sailcraft placed on the height of the synchronous orbit can maintain the fixed position with constant attitudes and lightness number.

For the hypothetical spherical asteroids, the perturbed gravitational accelerations from other celestial bodies are also ignored in this study. In fact, they may play a key role in the stability of the hovering orbit. Moreover, the variation of the Sun-asteroid distance due to the asteroids' eccentric orbits should have a great influence on the dynamics of the hovering orbit (Farres & Jorba 2012). All above effects should be taken into account in future studies. The stability of these hovering orbits and their controllability will be the subject of next publishing work. In terms of diverse shapes of the asteroids, a hovering study around elongated asteroids is in progress based on the current framework.

## 4 CONCLUSIONS

Feasible regions of asteroid body-fixed hovering have been investigated by using solar sailing. Four different sail models with the variable sail area are considered including the ideal, optical, parametric and SPT sails. Nonlinear equations are constructed to obtain the feasible hovering region and solved with a shooting method. With advanced thrust ability, the SPT sail can produce the largest hovering region which shrinks to smaller ones for other sails. For the boundary hovering radius of both inward and outward cases, the effect of the nonideal models must be considered in the mission design. The asteroid spinning rate plays a key role on the hovering region. With same physical and orbital characteristics, an asteroid with a higher value of rotating period (here is 15 hours corresponding to a slow rotating asteroid) holds a larger hovering region than the one with shorter period (9 hours). For a desired feasible hovering position away from the synchronous orbit, both the sail lightness number and its attitude have to be adjusted to counterbalance the asteroid rotation and the levitated gravitational acceleration out of the equatorial plane. For the hovering positions on the height of the synchronous orbit out of the equatorial plane, the sail acceleration is only provided to counterbalance the levitated gravitational acceleration resulting in the constant sail attitude and lightness number. Such a property is very attractive for solar sail body-fixed hovering missions and will be extended to different shaped asteroids in further studies.

**Acknowledgments**

This work was supported by the National Basic Research Program of China (973 Program, 2012CB720000) and Tsinghua University Initiative Scientific Research Program (No. 20131089268). Thank Dr. Jin-guang Li for his help about the Latex.